# X-band microwave generation caused by plasma-sheath instability

*Y. Bliokh, J. Felsteiner and Ya.Z. Slutsker*

Department of Physics, Technion, 32000 Haifa, Israel

**Abstract**

It is well known that oscillations at the electron plasma frequency may appear due to instability of the plasma sheath near a positively biased electrode immersed in plasma. This instability is caused by transit-time effects when electrons, collected by this electrode, pass through the sheath. Such oscillations appear as low-power short spikes due to additional ionization of a neutral gas in the electrode vicinity. Herein we present first results obtained when the additional ionization was eliminated. We succeeded to prolong the oscillations during the whole time a positive bias was applied to the electrode. These oscillations could be obtained at much higher frequency than previously reported (tens of GHz compared to few hundreds of MHz) and power of tens of mW. These results in combination with presented theoretical estimations may be useful, e.g., for plasma diagnostics.

**Introduction**

Instability of the plasma sheath around a positively-biased plasma-immersed electrode has been a known phenomenon during the last few decades [1]. This instability appears due to transit-time effects, when the collected electrons pass through the sheath. As a consequence, this instability causes oscillations in the electrode circuit with a frequency close to the electron plasma frequency $f_p$. These oscillations were of low power and could be detected either with the heterodyne method [1] or with a 30 dB amplifier before the detector [2]. All these experiments were performed in low dense (~ $10^9$ cm$^{-3}$) weakly ionized (~ $3 \cdot 10^{-2}$ %) plasma of a hot-cathode discharge. Values of $f_p$ typically did not exceed a few hundred megahertz, with maximum at ~ 1 GHz [1]. Also it should be pointed out that these oscillations appeared as periodic narrow spikes, of certainly less than 1 μs. These spikes were always correlated with strong low frequency (~ 100 kHz) oscillations of the electron current collected by this electrode (probe). These low frequency oscillations were caused by additional ionization of the neutral gas in the probe vicinity with "fireball" creation [2, 3]. The microwave (UHF) spikes at $f_p$ existed just together with the low frequency current oscillations. This is reasonable because the transit-time effects and the additional ionization appeared approximately at the same potential fall ~ 50 V between plasma and probe [1, 3]. The fireball effectively increased the electron-collecting area to achieve the current balance through the plasma [4]

In this paper we present first results of our study of the microwave spikes shortening, and the improvement of the UHF generation, its duration, frequency and power.

**Experimental setup**

The experiments were performed in a stainless steel vacuum chamber, having an inner diameter of 30 cm and a height of 8 cm (Fig. 1a). This chamber was equipped with two insulated tungsten filaments, one of them had 0.3 mm diameter and the other one 0.1 mm. Both of them could be heated separately. The chamber was also equipped with a large surface probe, having an area S = 7 cm$^2$ and with replaceable platinum probes of lengths 3 mm and diameters 0.05, 0.1 and 1.1 mm. The thicker tungsten filament was used as a hot cathode. It was heated when a voltage pulse $U_{heat}$ of 0.2 s duration was applied every 5 s. To obtain a hot-cathode discharge and a plasma we applied a dc voltage either between the cathode and the grounded vacuum chamber or between the cathode and the large surface probe. In the latter case the discharge circuit was floating. The plasma density $n_p$ and the electron temperature $T_e$ were derived from the C-V characteristics of the above-mentioned probes. Also we could use the platinum probe as a resonance probe [7]. To measure the plasma potential $U_h$ with respect to the chamber wall, the thin filament was used as a hot probe. To heat it, another dc source (small cell) was used. To measure the hot probe potential we used a divider with low input capacitance. To bias the small platinum probe we used either a dc power supply or a rectangular voltage with pulse duration 60 µs. Just a positive bias was used. The pulsed regime was used to prevent probe overheating if the probe collected current $I_{pr}$ was too high.

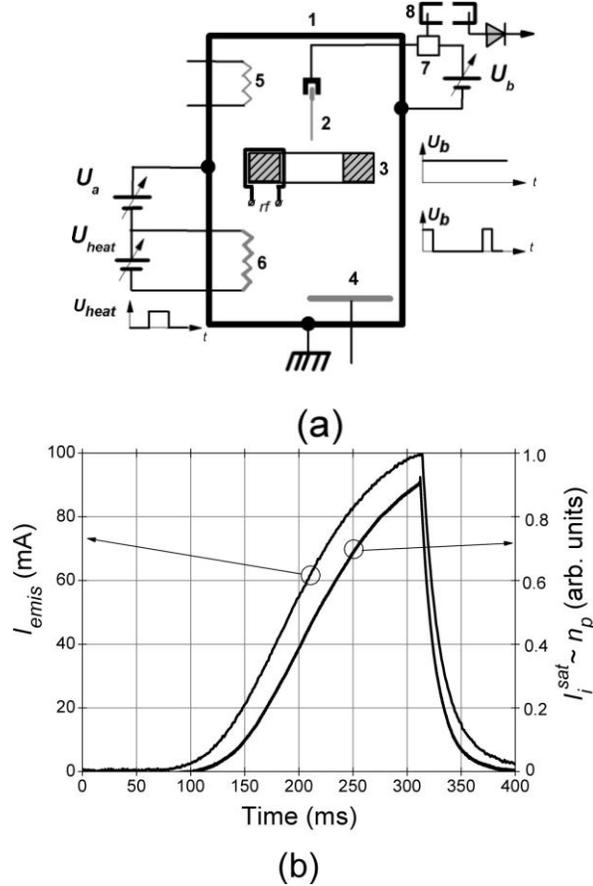

(a)

(b)

Fig.1. (a) 1 – vacuum chamber; 2 – replacable platinum probe, $U_b$ – bias voltage, either dc or rectangular pulses with 50 µs duration; 3 – ferromagnetic core of the FIC plasma source, inner diameter 10 cm, outer diameter 15 cm, thickness 2.5 cm, 15 turns of winding; 4 – large planar probe ($S = 7$ cm$^2$); 5 – thin tungsten filament (hot probe); 6 – hot cathode, $U_{heat}$ pulse duration 0.2 s, repetition rate 5 s; 7 – dc injector; 8 – combination of insulated coaxial waveguide adaptors. (b) left – cathode emitted current $I_{emis}$, $U_{heat} = 10.5$ V, $U_a = 100$ V, Xe gas; right – ion saturation current $I_i^{sat}$, collected by planar probe, $I_i^{sat} \sim n_p$.

In our experiments we used either Ar at a pressure (0.5 – 2) mTorr or (mainly) Xe at a pressure (0.15 – 1.1) mTorr. Typically we applied an anode voltage $U_a \sim 100$ V in order to make sure that we work in a regime with the thermal limitation of the emitted current. This is reasonable in order to eliminate dependence of the plasma density $n_p$ on $U_a$. Indeed in the range $U_a = (80 - 100)$ V, variations of the emitted current $I_{emis}$ did not exceed 10%. In the investigated pressure range the plasma density $n_p$ was found to be directly proportional to the pressure and the emitted current $I_{emis}$. The waveform of the emitted current $I_{emis}$ along with the ion saturation current $I_i^{sat}$, collected by the large surface probe, are shown in Fig. 1b. The duration of the $I_{emis}$ growth (~ 0.1 s) was long enough to consider a steady state plasma at any moment. In our experiments a 100 mA emission current corresponded to $n_p \leq 3 \cdot 10^9$ cm$^{-3}$, P = $3 \cdot 10^{-4}$ mTorr and $T_e \sim (2 - 4)$ eV with Xe gas. For the $n_p$ measurements the discrepancy between the probe characteristics method and the resonance probe method never exceeded 10 %. The fraction of fast electrons having energy $\sim eU_a$ never exceeded (1 – 2)%. They appeared as a step in the ion part of the probe characteristics. The emitted current $I_{emis}$ as well as the probe current $I_{pr}$ were measured with small current-view resistors. In order to study the influence of the ionization level we inserted inside the vacuum chamber (see Fig. 1a) a single-core ferromagnetic inductively-coupled (FIC) plasma source, driven by a powerful pulsed rf oscillator (see [5] and references therein). This oscillator could deliver up to 10 KW rf power which corresponded to rf voltage in the FIC entrance $U_{ind}$ = 450V, the rf pulse duration was about 2 ms. In these experiments the hot-cathode discharge was used as an initial electron supply. With this FIC source the plasma density increased a few hundred times (at the same pressure) and the ionization rate could reach a value of (20 – 30) %. It has been shown recently that in this case there is no ionization instability in the probe vicinity [6]. Also the electron temperature $T_e$ was found to be (2 – 4) eV [5]. The thin platinum probe was connected to the bias power supply via a dc injector, which enabled us to separate the dc and UHF signal. For the low-dense plasma obtained with the hot-cathode discharge, we used LC filters with cut-off frequency of 60 MHz. It worked properly up to 500 MHz. For the microwave signals we used a hand-made dc injector, based on coaxial-waveguide adaptors (CWA). For the microwave measurements we used one CWA for (8.2 – 12.4) GHz as a dc injector and three wide-band waveguide detectors for (8.2 – 12.4) GHz, (12.2 – 18) GHz and (18 – 26.5) GHz, the last two with corresponding waveguide adaptors for (8.2 – 12.4) GHz waveguide (Fig. 1a). Also we used a tunable detector for (8.2 – 12.4) GHz. This detector had a bandwidth of (300 – 400) MHz (depending on the central frequency). All these elements could be connected to each other via insulators to prevent parasite signals via ground loops.

**Experimental results**

When a positive bias above ~ 60 V was applied to the small probe immersed in the hot-cathode discharge plasma, very typical current oscillations $I_{pr}$ appeared in the probe circuit (Fig. 2a). These oscillations manifested the plasma-sheath instability caused by local additional ionization in the probe vicinity [3]. A further increase of $U_b$ caused appearance of narrow microwave spikes whose shape and duration is very similar to those presented in Refs. [1,2] (see Fig. 2b). On the other hand, the amplitude of the UHF oscillations could reach 0.3 V (~ 1 mW) at 50 Ω loading, which is a few orders of magnitude higher than in Refs. [1,2]. As it is clearly seen from the expanded spike, the UHF frequency is approximately 290 MHz (see Fig. 2c) which is in good agreement with the measured plasma density $n_p \sim 10^9$ cm$^{-3}$. To be sure, we used the same electrode as a resonance probe [7] and obtained the maximal reflected signal at the same frequency ~ 290 MHz (Fig. 2d). Such UHF spikes appeared in the bias range (80 – 150) V which is a bit higher compared to [1, 2].

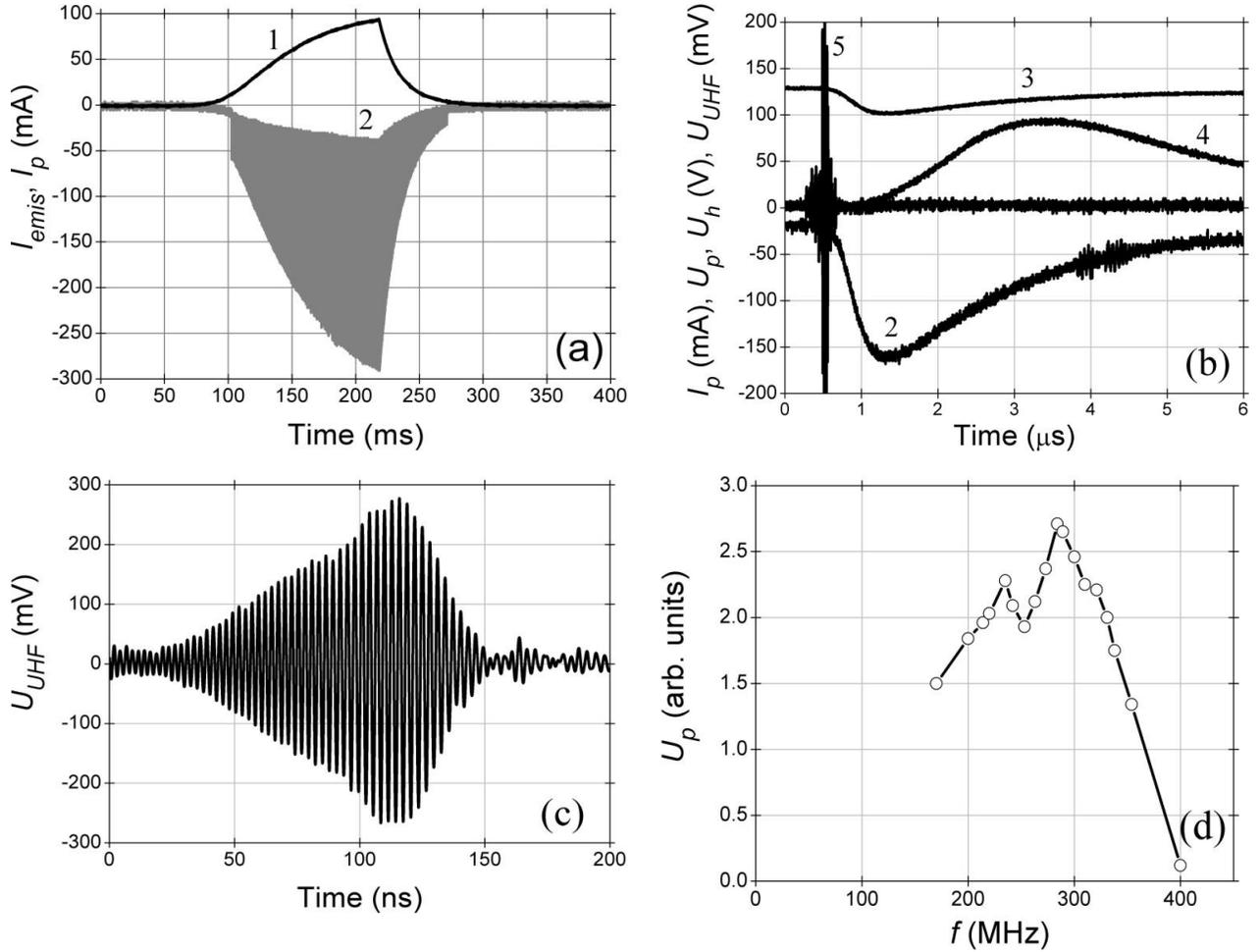

Fig.2. (a) 1 – cathode emitted current $I_{emis}$; 2 – electron current $I_{pr}$, collected by the thin (0.05 mm) probe. (b) expanded waveforms near $I_{emis} \approx 40$ mA ($n_p \sim 10^9$ cm$^{-3}$); 2 – electron current $I_{pr}$; 3 – bias voltage $U_b$; 4 – plasma potential with respect to the chamber walls; 5 – UHF spike in the probe current $I_{pr}$. (c) expanded UHF spike, $f_p \approx 290$ MHz. (d) the same probe, used as a resonance probe, maximal reflected signal at the same frequency.

On the other hand in these experiments we were not able to obtain CW UHF generation or even to make the spikes wider. This is clear because the UHF instability caused by transit-time effects exceeds the bias voltage $U_b$, required for ionization instability in the probe vicinity. To prevent the additional ionization in the probe vicinity we increased drastically the plasma density at the same pressure, i.e. the level of ionization of the gas [6]. As it was mentioned above, we used a pulsed (2 ms) FIC plasma source [5] (Fig. 1a) which produced a pulse of high-dense plasma (Fig. 3a).

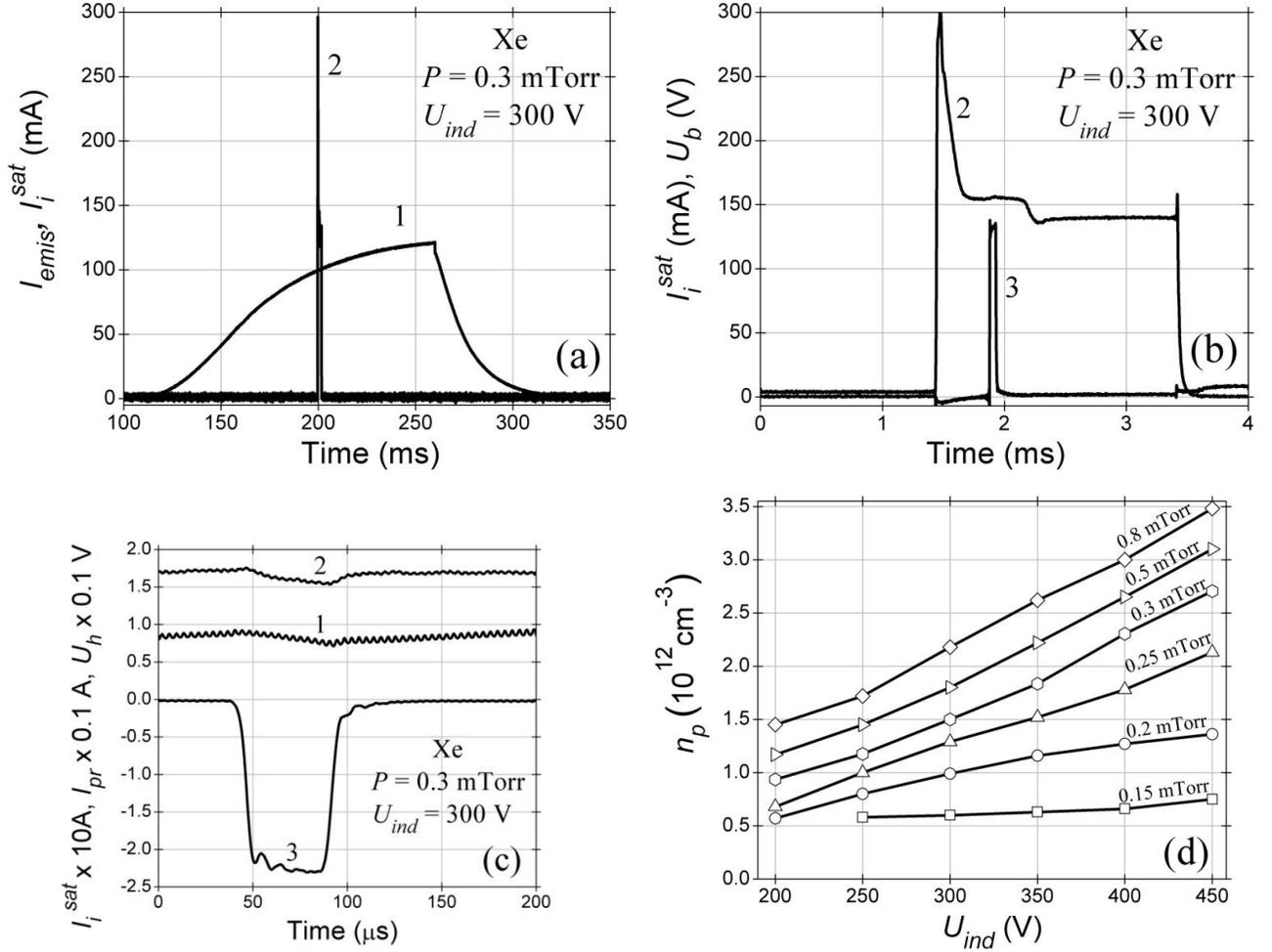

Fig.3. (a) 1 – emitted current $I_{emis}$; 2 – ion saturation current $I_i^{sat}$, collected by the planar probe during FIC rf driving pulse. (b) 2 – expanded $I_i^{sat}$ pulse; 3 – bias pulse $U_b$. (c) 3 – electron current $I_{pr}$, collected by the thin (0.05 mm) probe. d: plasma density $n_p$ vs $U_{ind}$ for various pressures of Xe gas. Derived from $I_i^{sat}$ to the large planar probe, $T_e \sim 4$ eV.

To prevent the probe damage when the plasma density was high, we applied the positive bias just during 50μs (Fig. 3b). Nevertheless this duration definitely exceeded the ionization time [2-4,6]. A typical detected UHF pulse obtained with combination of two 3-cm (8.2 – 12.4 GHz) waveguides and a wide-band detector is shown in Fig.4a. This waveform corresponds to undisturbed quasi-neutral plasma density ~1.5·10$^{12}$ cm$^{-3}$, i.e. with plasma resonance frequency $f_p \sim 12$ GHz. A similar waveform of microwave signal was obtained with Ar (Fig. 4b). Note, that though the gas pressures in both cases were very different, the voltage needed to obtain microwave oscillations was approximately the same. The estimated plasma densities via electron temperature and ion saturation current to the big surface probe were also close. The microwave waveform and its amplitude definitely depended on the probe diameter. So, we did not see any difference when we used probes of 50 and 100 μ, but when we moved to the probe diameter 1.1 mm, the waveforms were changed drastically: they appeared as spikes, mainly connected to the beginning and the end of the bias pulse (Fig. 4c). They appeared just in a narrow gas pressure range (0.3 – 0.35 mTorr of Xe) and a narrow range of driving rf power (rf voltage across the FIC was 360 – 370 V). This 1.1-mm probe in the dense plasma ($n_e \sim 10^{12}$ cm$^{-3}$) may be recognized as a planar probe ($r_d \sim 10^{-3}$ cm at $T_e \sim 3$ eV). So, one may conclude that in this case the microwaves appeared just during the sheath formation and its decay when the bias pulse $U_b$ is applied. In both these cases the ratio of the small probe area to the total chamber wall area $S_p/S$ was very small and significantly less than $\sqrt{m/M}$ even for Xe gas (here $M$ and $m$ are the electron and ion mass respectively). This means that almost all the bias voltage is applied between the small probe and the plasma because $U_h$ is small (Fig. 3c). In these measurements, when the plasma was dense, we did not see periodic low frequency probe current spikes related to additional ionization in the probe vicinity as shown in Fig. 2a (for more details see [3,4] and references therein). When the electron current exceeded a certain level (30 – 50) A (depending on the probe bias, pressure, etc.), the thin probe could be damaged (evaporated). This was an actual limit of the present measurements. When the plasma density became higher because of higher pressure or/and rf power, the threshold value of $U_b$ became smaller.

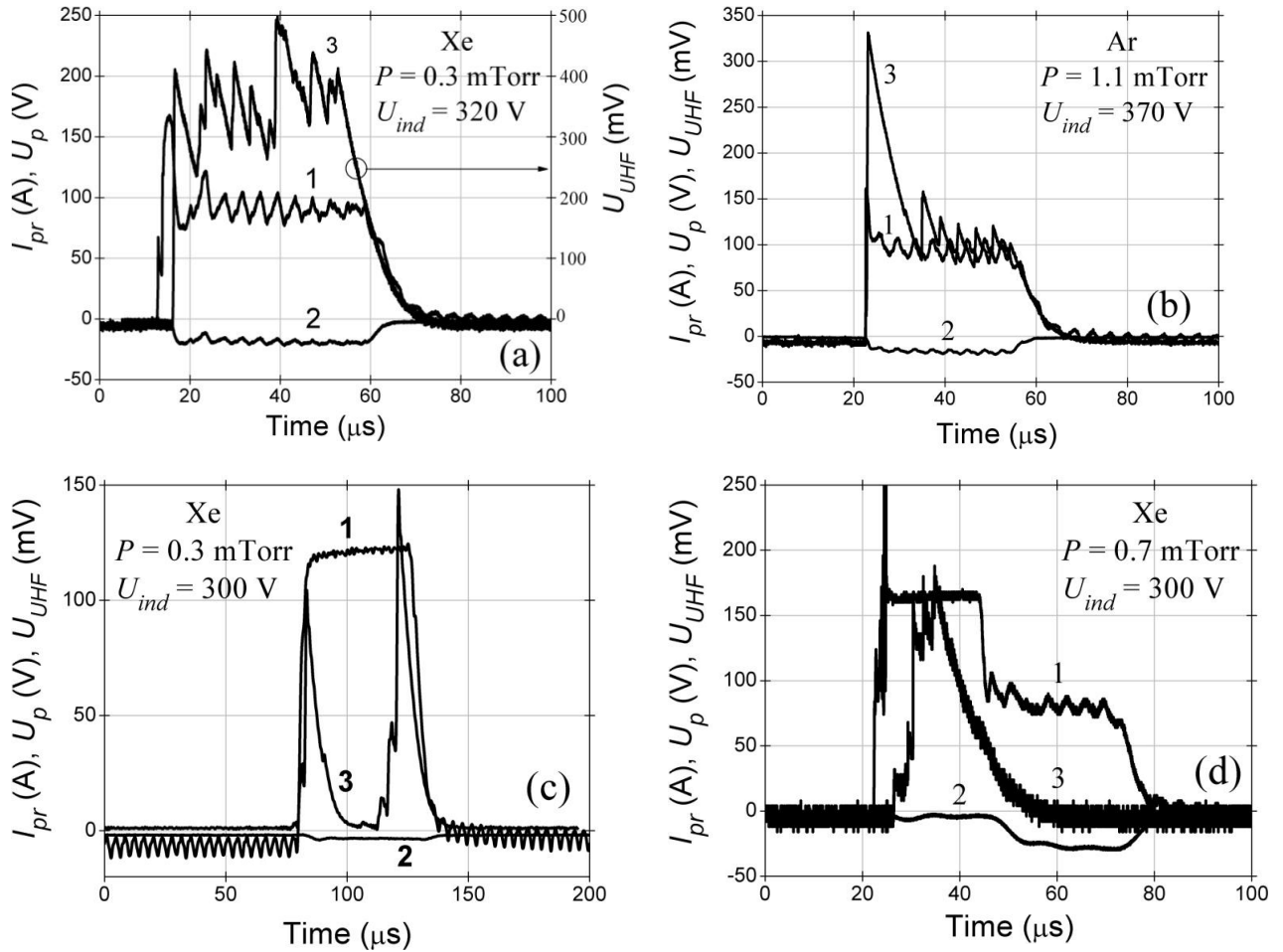

Fig.4. (a) Xe gas. (b) Ar gas. 1 – bias voltage $U_b$; 2 – electron current $I_{pr}$, collected by the thin (0.05 mm) probe; 3 – detected UHF signal. (c) the same, but the probe diameter is 1.1 mm. (d) the thin (0.05 mm) probe, but the current $I_{pr}$ is high (25 A).

The microwave signal shape and its maximal value did not repeat each other in all details from pulse to pulse. The amplitude difference could achieve 1.5 times. Nevertheless it is possible to state that microwave generation appeared in the probe current range (3 – 15) A. At higher $I_{pr}$ there was no microwave generation even if the probe was not damaged (Fig. 4d). The maximal detected microwave signal definitely depended on the gas pressure. At low pressures (0.15 mTorr) it could exceed 1 V while at high pressures (1 mTorr) it rather exceeded 100 mV (Fig. 5a). The probe current $I_{pr}$ which corresponds to the maximal microwave signal tended to be smaller with the pressure rise (Fig. 5a). The rf driving voltage in these measurements usually could be found in the range (350 –

400) V (higher at lower pressure). The microwave generation occured in a wide range of rf driving powers and gas pressures. On the other hand the generation appeared just in a comparatively narrow range of the bias voltage $U_b$. Thus, the voltages $U_b$, when the generation takes place for various $U_{ind}$ (various $n_p$) and $P$ is presented in Fig. 5b. It is clearly seen there that the microwave generation occurs just when 90 V < $U_b$ < 150 V. The dependence of the microwave signal $U_{UHF}$ on $U_{ind}$ for various pressures $P$ is presented in Fig. 5c. It is seen there that for low pressures there is a certain plasma density (certain $U_{ind}$), below which the microwave signal disappears. As it is seen in Fig. 3d, a pressure of 0.25 mTorr and $U_{ind} \leq 200$ V correspond to $n_p \sim 5.5 \cdot 10^{11}$ cm$^{-3}$ and consequently $f_p \sim$ 6.5 GHz, which is the cut-off frequency of the waveguide we used (2.3 cm width). For high gas pressure (0.8 mTorr) there is a certain $U_{ind}$ above which the signal also disappears. As seen in Fig. 3d, a pressure of 0.8 mTorr and $U_{ind} = 330$ V correspond to $n_p \sim 2.5 \cdot 10^{12}$ cm$^{-3}$ and, therefore, $f_p \sim 14$ GHz or a wavelength of 2.1 cm. The half of this value corresponds to the height of the waveguide we used (1.05 cm), i.e. an effective TE- mode transformation becomes possible. Other points corresponding to these thresholds are shown in Fig. 5d. All of them correspond approximately to the same plasma densities (see Fig. 3d). For a crude estimation of the microwave spectrum width the detector for (8.2 – 12.4) GHz could be replaced by detectors for frequency (12.4 – 18) GHz or (18 – 26.5) GHz. As we already mentioned these detectors could be connected to the probe-connected CWA (8.2 – 12.4) GHz via waveguide adaptors. The microwave signal obtained with the (12.4 – 18) GHz waveguide was one order of magnitude smaller than that with (8.2 – 12.4) GHz, and with the (18 – 26.5) GHz waveguide there was no detected signal at all. When we used a narrow band microwave detector, which could be tuned in the frequency range (8.2 – 12.4) GHz, the microwave signal never filled all the bias pulse (see Fig. 4a,b), but always appeared as a narrow spike, which could be found inside the pulse of bias voltage $U_b$ (Fig. 6). From pulse to pulse this microwave spike could be shifted inside the bias pulse and change its magnitude by (1.5 – 2) times. Comparing signals from the tunable detector and the wide-band detector one may conclude that in case of dense plasma the radiated spectrum is wide, each component appears at unpredictable moment during the pulse, i.e. the signal is randomly frequency-modulated.

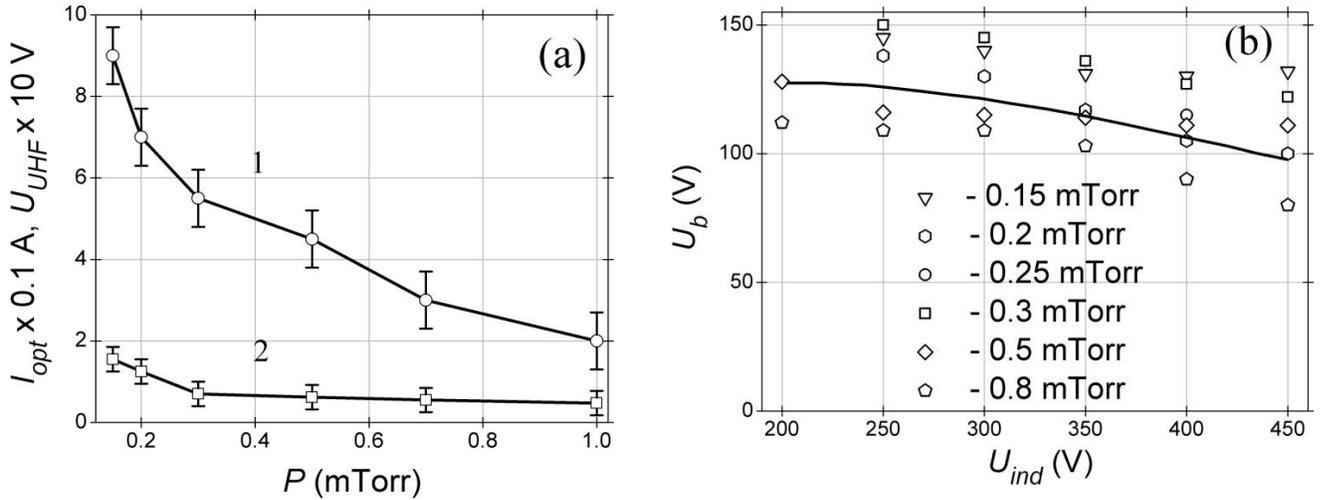

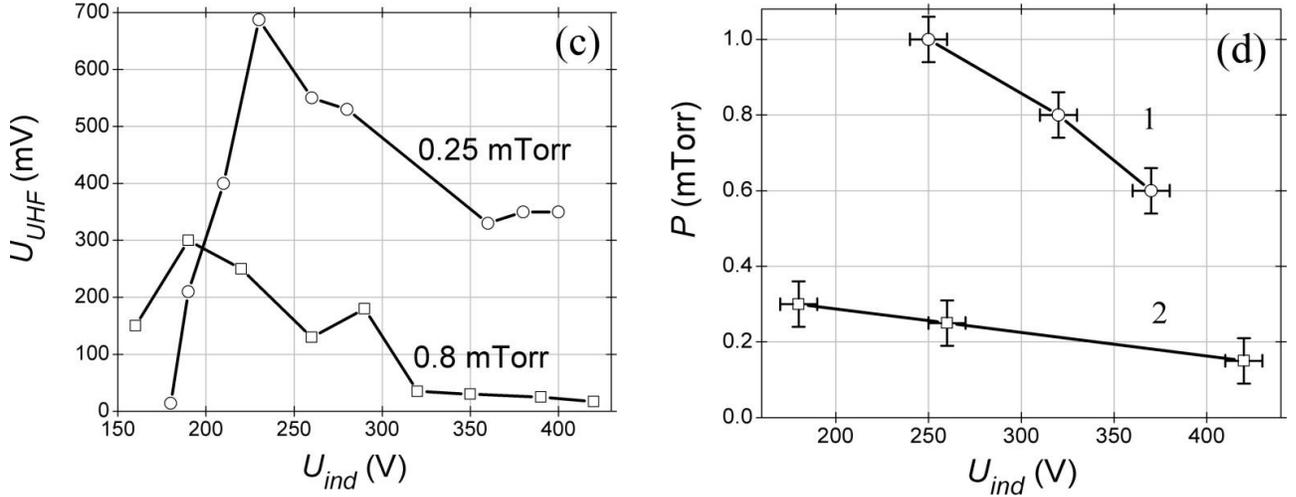

Fig.5. Xe gas. (a) 1 – microwave signal $U_{UHF}$ vs gas pressure $P$; 2 – optimal probe current $I_{pr}$, $U_{ind} \sim (350 - 400)$ V. (b) bias voltage $U_b$ vs $U_{ind}$ for various pressures $P$. (c) microwave signal $U_{UHF}$ vs $U_{ind}$ for various pressures $P$. (d) "breaking points" of microwave signal for low and high pressures.

It is interesting to note, that in our study with low-dense plasma the UHF generation started at higher voltage than in Ref. [1].This is understood because in [1] the study was done during plasma decay when the electron temperature was low. Moreover, in our study the range of $U_b$ was the same for low and high dense plasmas and the electron temperature was approximately the same.

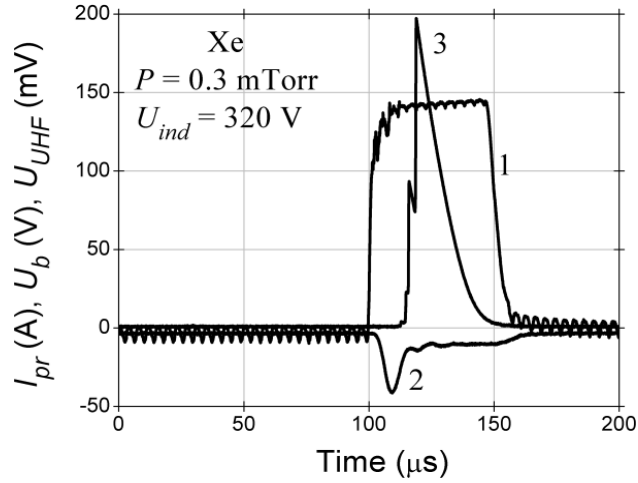

Fig.6. Microwave signal, obtained with narrow-band detector: 1 – bias voltage $U_b$. 2 – probe current $I_{pr}$. 3 – microwave signal $U_{UHF}$.

**Theoretical model**

As it was mentioned by Stenzel in Ref. [1], the resonant response of the ambient plasma on the oscillating space charge inside the sheath around the positively biased electrode can make the sheath unstable. The possible reason of this resonant plasma response is the following. Plasma Langmuir oscillations near the sheath boundary modulate the flux of plasma electrons through the sheath boundary and, as a consequence, result in temporal variations of the total space charge inside the sheath volume. The oscillating total charge of the sheath produces electric field in the vicinity of the sheath boundary. When this field matches with the field of plasma oscillations, the oscillations amplitude can grow exponentially in time.

This qualitative consideration allows formulation of the following mathematical model. Let us consider the simplest step-like model of a steady-state electron-rich plasma sheath of radius $r_{sh}$ around a small positively biased cylindrical electrode. A quasi-neutral plasma occupies the region $r > r_{sh}$. The flux of thermal electrons from the plasma into the sheath region creates a space charge that shields the electrode potential so that the electric field $E_{sh}$ at the plasma boundary $r = r_{sh}$ is equal to zero, $E_{sh} = 0$. The temporal variation of the electron flux produces an unbalance $\tilde{q}(t)$ of the total space charge inside the sheath volume and an uncompensated electric field $\tilde{E}_{sh}(t) = 2\tilde{q}(t)/r_{sh}$ at the sheath boundary. It is worth noting that such a simple relation between the electric field $\tilde{E}_{sh}(t)$ and the charge $\tilde{q}(t)$ is valid in the cylindrically symmetric configuration (a similar linear relation between $\tilde{E}_{sh}(t)$ and $\tilde{q}(t)$ is valid in a spherically symmetric configuration also). In a planar configuration (flat electrode) this relation is more complicated, namely, the electric field depends not only on the charge $\tilde{q}(t)$ but also on the charge spatial distribution $\tilde{q}(t,z)$ in the sheath as well (see, e.g., [8]).

The charge $\tilde{q}(t)$ can be expressed in terms of the electron radial velocity variation $\tilde{v}(t)$ at the sheath boundary as follows:

$$\tilde{q}(t) = -2\pi e r_{sh} \frac{n_p}{2} \int_{t-T}^{t} dt'\, \tilde{v}(t'), \qquad (1)$$

where $T$ is the electron transit time through the sheath. It is taken into account in Eq (1) that only electrons with a negative unperturbed radial velocity penetrate into the sheath region, and the density of these electrons is $n_p/2$. Now, one can write the equation of electron motion at the sheath edge:

$$\frac{d\tilde{v}}{dt} = \frac{e}{m}\tilde{E}_{sh}(t) = -\frac{\omega_p^2}{2}\int_{t-T}^{t} dt'\,\tilde{v}(t'). \qquad (2)$$

Eq. (2) admits a monochromatic solution $\tilde{v}(t) \sim e^{-i\omega t}$, where the frequency $\omega$ is the root of the following dispersion equation:

$$\omega^2 = -\frac{\omega_p^2}{2}\left(1 - e^{i\omega T}\right). \qquad (3)$$

As it was argued above, the sheath can be unstable when the field $\tilde{E}_{sh}$ is matched with the field of the plasma waves, i.e. the frequency $\omega$ is close to the plasma Langmuir frequency $\omega_p$. The dispersion equation (3) has a solution $\omega = \omega_p + \delta\omega$, $|\delta\omega| \ll 1$, when $\omega_p T = (2k+1)\pi + \delta\alpha$, $|\delta\alpha| \ll 1$. In this case Eq. (3) can be written as:

$$\delta\omega = \frac{i}{4}\omega_p \delta\alpha. \qquad (4)$$

Thus, the sheath is unstable if the transit angle (dimensionless transit time) $\theta \equiv \omega_p T = (2k+1)\pi + \delta\alpha$ and $\delta\alpha > 0$. The transit angle value $\theta = \theta_{th} = \pi$ can be treated as the instability threshold. Note, that in the flat configuration the threshold value of the transit angle is equal to $2\pi$ [1,8]. This is associated with the difference between the flat and cylindrical configurations mentioned above.

In order to find the range of parameters where the sheath is unstable, it is necessary to determine the potential distribution $\varphi(r)$ in the steady-state sheath and define the transit time $T$ as a function of these parameters. The potential is defined by simultaneous solution of the Poisson equation, the continuity equation, and the equation of electron motion. In the experimental conditions the electrode radius $r_e = 70\,\mu$ is large as compared with the Debye radius $r_d$ (the relation $r_e/r_d$ is varied in the range $4.9 \leq r_e/r_d \leq 10.2$ under the plasma density variation from $0.8 \cdot 10^{12}\,\text{cm}^{-3}$ to $3.5 \cdot 10^{12}\,\text{cm}^{-3}$). Taking the latter into account, one can neglect the azimuthal motion of the plasma electrons in the sheath region and consider the one-dimensional model of the cylindrically symmetric sheath as a rough approximation. It is assumed in this model that electrons incoming into the sheath have an initial radial velocity $v_r = -v_{Te}$, where $v_{Te}$ is the thermal velocity. Omitting obvious calculations, the resulting equation that describes the electric potential distribution $\varphi(r)$ in the cylindrical sheath can be written in the following dimensionless form:

$$\frac{d^2\psi}{d\rho^2} = -\frac{1}{\rho}\frac{d\psi}{d\rho} + \frac{\rho_{sh}}{\rho}\frac{1}{\sqrt{1+\psi}}, \qquad (5)$$

where $\psi = -2e\varphi/mv_{Te}^2$ is the dimensionless potential, $\rho = r/r_d$ is the dimensionless radius, $r_d = v_{Te}/\omega_p$ is the Debye radius, and $\rho_{sh} = r_{sh}/r_d$. The boundary conditions are $\psi(\rho_{sh}) = d\psi/d\rho\big|_{\rho=\rho_{sh}} = 0$ (the last condition defines the sheath radius $\rho_{sh}$) and $\psi(\rho_e) = \psi_e$, where $\rho_e$ is the electrode dimensionless radius, and $\psi_e$ is the electrode dimensionless potential. The transit angle $\theta = \omega_p T$ is defined as $\theta = \int_{\rho_{sh}}^{\rho_e} d\rho (1+\psi)^{-1/2}$.

As it was mentioned above in the previous section, the UHF signal appears when the probe potential exceeds a certain threshold value $\varphi_{th}$. This threshold potential depends on experimental conditions, as it is shown in Fig. 5b. A numerical solution of Eq. (5) with parameters which are in accordance with the experimental data presented in Fig. 3d and 5b, allows one to calculate the transit angles $\theta_{th}$ associated with $\varphi_{th}$. Results of the calculations, which were carried out for pressures 0.8 mTorr and 0.2 mTorr, are shown in Fig. 7. Regardless of the fact that the plasma density values, which correspond to these experimental data, vary in a broad range (from $0.8 \cdot 10^{12}\,\text{cm}^{-3}$ to $3.5 \cdot 10^{12}\,\text{cm}^{-3}$), the transit angles $\theta_{th}$ are practically the same for all experimental points and exceed the value $\theta_{th} = \pi$ predicted by the theoretical model.

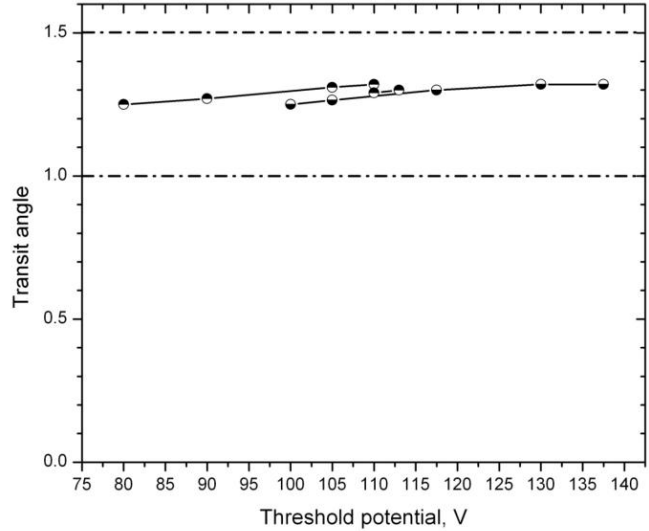

Fig. 7. Theoretical values of the transit angles (in $\pi$ units) which correspond to the experimentally measured threshold potentials $\varphi_{th}$. Top-filled circles – pressure 0.8 mTorr, bottom-filled circles - pressure 0.2 mTorr. All other experimental points which correspond to pressures between 0.2 mTorr and 0.8 mTorr, are located between the depicted curves.

It is interesting to note that in our study the optimized probe size for low-dense plasmas ( $d = 1.1$ mm at $n_p = 3 \cdot 10^9 - 10^9 \, \text{cm}^{-3}$ ) corresponds to the same values of parameters $\rho_e$ and $\rho_{sh}$, and the instability appears at the same bias $U_b$. This means that the transit angle $\theta_{th}$ is also the same in case of low-dense plasmas.

**Conclusion**

In the present study of plasma-sheath instability near an electron-collected electrode we succeeded to obtain in the electrode circuit a much longer microwave pulse at much higher microwave power and frequency. The suggested theory qualitatively explains the UHF instability existence. The fact that the oscillation frequency is close to the plasma frequency makes this simple device promising for plasma diagnostics. Indeed, there is no direct influence of the electron temperature like in Langmuir probe measurements, the electrode could be very small, so it does not disturb the plasma. This method does not require an external microwave generator like in the cut-off or resonance-probe methods. Also the bias pulse may be quite short, just enough to measure the oscillation frequency. On the other hand, the generated microwave signal did not have a stable amplitude and frequency. This could be caused by various nonlinear processes in the electrode vicinity which is a subject for further investigations as well as for a more detailed study of the microwave spectrum.

**List of References**